# Large influence of capping layers on tunnel magnetoresistance in magnetic tunnel junctions

Jiaqi Zhou,[1,2,3] Weisheng Zhao,[1,2,a)] Yin Wang,[4] Shouzhong Peng,[1,2] Junfeng Qiao,[1,2] Li Su,[1,2,3] Lang Zeng,[1,2] Na Lei,[1,2] Lei Liu,[5] Youguang Zhang,[1,2] and Arnaud Bournel[3]

[1]*Fert Beijing Institute, Beihang University, BDBC, Beijing, 100191, China*

[2]*School of Electronic and Information Engineering, Beihang University, Beijing, 100191, China*

[3]*Centre for Nanoscience and Nanotechnology, Univ. Paris-Sud, Université Paris-Saclay, CNRS, F-91405 Orsay, France*

[4]*Department of Physics and the Center of Theoretical and Computational Physics, The University of Hong Kong, Hong Kong, China*

[5]*Nanoacademic Technologies Inc., Brossard, QC J4Z 1A7, Canada*

It has been reported in experiments that capping layers which enhance the perpendicular magnetic anisotropy (PMA) of magnetic tunnel junctions (MTJs) induce great impact on the tunnel magnetoresistance (TMR). To explore the essential influence caused by capping layers, we carry out *ab initio* calculations on TMR in the X(001)|CoFe(001)|MgO(001)|CoFe(001)|X(001) MTJ, where X represents the capping layer material which can be tungsten, tantalum or hafnium. We report TMR in different MTJs and demonstrate that tungsten is an ideal candidate for a giant TMR ratio. The transmission spectrum in Brillouin zone is presented. It can be seen that in the parallel condition of MTJ, sharp transmission peaks appear in the minority-spin channel. This phenomenon is attributed to the resonant tunnel transmission effect and we explained it by the layer-resolved density of states (DOS). In order to explore transport properties in MTJs, the density of scattering states (DOSS) was studied from the point of band symmetry. It has been found that CoFe|tungsten interface blocks scattering states transmission in the anti-parallel condition. This work reports TMR and transport properties in MTJs with different capping layers, and proves that tungsten is a proper capping layer material, which would benefit the design and optimization of MTJs.

Since the discovery of tunnel magnetoresistance (TMR) in the Fe|Ge-oxide|Co trilayer structure,[1] the TMR effect has become the significant principle of non-volatile magnetic random access memory (MRAM).[2] As the basic element of MRAM, magnetic tunnel junction (MTJ) was investigated in detail.[3] Thanks to rapid advances in growth technique of ultra-thin ferromagnetic films, CoFeB|MgO MTJs with high TMR ratio have been achieved[4-7] as predicted by *ab initio* calculations.[8] Recently, based on MTJs with high TMR and the spin transfer torque effect, the spin-transfer-torque MRAM (STT-MRAM) is achieved with merits of high density and low power consumption.[9]

---

a) Electronic mail: weisheng.zhao@buaa.edu.cn

As the *ab initio* theory successfully explained the high TMR of MTJ,[8] a great deal of progresses have been achieved by *ab initio* calculations. Butler *et al.* provided a thorough explanation of the physics behind the spin-dependent tunnel conductance of Fe|MgO|Fe from the aspect of symmetry matching of Bloch states.[10] Zhang *et al.* reported a TMR investigation in body-centered cubic Co|MgO|Co and FeCo|MgO|FeCo tunnel junctions, indicating that the total reflection depends on the absence of the $\Delta_1$ band for the minority spin in the cubic direction, and predicted a high TMR ratio in the MTJ based on FeCo.[11] Waldron *et al.* researched the voltage dependence of TMR in the Fe|MgO|Fe MTJ at non-equilibrium, suggesting that the quench of TMR by bias was due to a relatively fast increase of channel currents in the anti-parallel condition.[12] Furthermore, plenty of studies have reported that subtle details at the metal|insulator interface could influence TMR and spin properties of the MTJ, such as vacancy defects at the interface, oxidation of the magnetic metal at the interface, presence of interfacial resonant states, etc.[13-15]

In addition to TMR, another important device trait of the MTJ is the perpendicular magnetic anisotropy (PMA), which is critical for achieving high thermal stability in spintronic devices. The PMAs observed in MgO|CoFeB|metallic capping layer structures have been investigated extensively[16-20] and it has been demonstrated that PMAs of CoFeB|MgO-based structures crucially depend on the capping layer material.[21-23] Recently, experiments have investigated the influence on TMR caused by capping layers, and drawn the conclusion that a higher TMR can be achieved in MTJs with W capping layers instead of Ta.[24-26] However, the understanding of the essential influence on TMR caused by capping layers remains unclear, leading to difficulties in TMR optimization for MTJ nano-pillars.

In this work, we report an *ab initio* study on the spin-dependent transport in X(001)|CoFe(001)|MgO(001)|CoFe(001)|X(001) MTJ nano-pillar, where X represents the capping layer material tungsten, tantalum or hafnium as strong PMA can be achieved in MgO|CoFeB|X structures.[16,20,23] The atomic structure is a two-probe MTJ that is divided into three regions: the left|right semi-infinite electrodes made of X, and the MTJ structure as scattering region, which is shown in Fig. 1. The MTJ structure consists of five X monolayers as cap or seed layer, five CoFe monolayers as ferromagnetic layer, and five MgO monolayers as barrier layer. The charge transport of the MTJ is along the z-direction, and the MTJ structure is periodically repeated along x- and y- directions. The x- and y-lattice constant of the junction is fixed to 2.83 Å.[21] The X|CoFe interfaces have been setup with the crystallographic orientation of X(001)[110]||CoFe(001)[100] to minimize the lattice mismatch. Atomic structures of central scattering region is relaxed by Vienna *ab initio* simulation package (VASP),[27] and the optimized distance from the interfacial X layer to the closest Co layer is 1.713 Å, 1.792Å and 1.837 Å when X represents W, Ta or Hf, respectively.

Quantum transport properties were calculated by a state-of-the-art technique based on density functional theory (DFT) combined with the Keldysh non-Equilibrium Green's function (NEGF) formalism[28,29] as implemented in the NanoDCAL package. In the NEGF-DFT transport simulation, the physical quantities are expanded by a linear combination of atomic orbital (LCAO) basis sets at the double-zeta plus polarization orbital (DZP) level. The spin-resolved conductance is obtained by the Landauer-Büttiker formula

$$G_\sigma = \frac{e^2}{h} \sum_{\mathbf{k}_\parallel} T_\sigma(\mathbf{k}_\parallel, E_F) \tag{1}$$

in which $T_\sigma(\mathbf{k}_\parallel, E_F)$ is the transmission coefficient with spin $\sigma$ at the transverse Bloch wave vector $\mathbf{k}_\parallel = (k_x, k_y)$ and the Fermi level $E_F$, $e$ is the electron charge and $h$ is the Planck constant. The spin-resolved transmission coefficient at $E_F$ is calculated by

$$T_\sigma(E_F) = \text{Tr}[\Gamma_L(E_F)G^r(E_F)\Gamma_R(E_F)G^a(E_F)]_{\sigma\sigma} \qquad (2)$$

where $G^r$ and $G^a$ are respectively the retarded and advanced Green's functions of the system. $\Gamma_\alpha$ ($\alpha$ = L, R) is the linewidth function which describes the coupling between the α lead and the scattering region. A 20×20×1 k-point mesh is sufficient for the NEGF-DFT self-consistent calculation of this two-probe device, and a much denser sampling of 300×300×1 was used for calculating the transmission coefficient in order to converge the summation in Eq. (1). The mesh cut-off energy was set to be 3000 eV. The energy tolerance for self-consistency was restricted to $10^{-5}$ eV. Using the calculated conductance, the TMR ratio is obtained as

$$TMR = \frac{G_{PC} - G_{APC}}{G_{APC}} \qquad (3)$$

where $G_{PC}$ and $G_{APC}$ are the total conductance for the magnetizations of two ferromagnetic layers in the parallel configuration (PC) and anti-parallel configuration (APC) respectively.

We investigated the spin-resolved transport properties in PC and APC of MTJs. The spin-resolved conductance is presented in Fig. 2 and normalized TMR ratios are shown in the inset. $G_{PC}^{\uparrow\uparrow}$ and $G_{PC}^{\downarrow\downarrow}$ is the majority-spin and minority-spin conductance in PC respectively. $G_{APC}^{\uparrow\downarrow}$ and $G_{APC}^{\downarrow\uparrow}$ are the majority-to-minority and minority-to-majority conductance in APC respectively with almost the same value. It can be seen that $G_{PC}^{\uparrow\uparrow}$ is larger than the conductance in APC, due to the $\Delta_1$ spin-filtering effect.[11] The TMR ratio of the W-capped MTJ is higher than that of Ta-capped and Hf-capped MTJs due to the low conductance in APC. This result corresponds to experimental results where the TMR in W-capped MTJ is much higher than that in Ta-capped MTJ.[24-26] In PC, by observing the values of $G_{PC}^{\uparrow\uparrow}$ and $G_{PC}^{\downarrow\downarrow}$, we find that $G_{PC}^{\uparrow\uparrow}$ in three MTJs are almost the same. However, $G_{PC}^{\downarrow\downarrow}$ in Hf-capped MTJ is much larger than that in W-capped and Ta-capped MTJs.

To understand the spin-resolved conductance and transport properties, we investigated transmission coefficients with log scale in the two-dimensional Brillouin zone (BZ) at $E_F$, as shown in Fig. 3. For all the three X-capped MTJs, the majority spin in PC has a broad peak centered at $\mathbf{k}_\parallel = (0,0)$ due to the slow decay of the $\Delta_1$ state, whereas for the minority spin in PC and APC there are negligible transmission probabilities, except some very sharp peaks at special $\mathbf{k}_\parallel$ points appear in PC (see bright spots in red circles). Fig. 3(h) shows that at $\mathbf{k}_\parallel = (0.12, 0.97)$ $\pi/a$, the channel has a transmission coefficient larger than 0.4, indicating that electrons transmit through the MgO tunnel barrier with over 40% probability. This channel significantly contributes to $G_{PC}^{\downarrow\downarrow}$ in the Hf-capped MTJ. Such a high transmission probability originates from resonant tunnel transport[30] and these $\mathbf{k}_\parallel$ points are called hot spots. It occurs when localized interfacial states align in energy, which can be confirmed by the layer-resolved partial density of states (DOS) of the minority spin in Hf-capped MTJ as shown in Fig. 4. Clear peaks appear at CoFe|MgO and CoFe|Hf interfaces, indicating the existence of resonant states. This resonance tunnel transmission also appears when X is W or Ta, as shown in red circles in Fig. 3(b) at $\mathbf{k}_\parallel = (0.80, 0.80)\pi/a$ with the transmission coefficient of 0.018 and Fig. 3(e) at $\mathbf{k}_\parallel = (0.99, 0.75)\pi/a$ with 0.008. Note that the tunnel transport is sensitive to device details. Any small variation in the MTJ device is enough to shift the narrow energy bands which reduces the desired TMR.[31]

In order to investigate transport properties in MTJs, we researched the density of scattering states (DOSS) in each atomic layer, which is shown in Fig. 5. DOSS presents the scattering states number in unit energy for transport. Majority- to majority-spin channel in PC condition and majority- to minority-spin channel in APC condition were analyzed. We focused on the band with $\Delta_1$ symmetry states as $\Delta_1$ component dominate tunnel in MTJs with the MgO barrier layer.[11] It can be found that in majority- to majority-spin condition, all scattering states behave similarly and terminate at around $10^{-5}$ orders of magnitude. However, an obvious difference appears at the outgoing CoFe|X interface in majority- to minority-spin condition. Scattering states in Hf-capped MTJ terminate at the $10^{-6}$ orders of magnitude, while that in W-capped MTJ terminate below $10^{-7}$ orders of magnitude. We attribute this phenomenon to the orbital hybridization caused by capping layers, and the hybridization changes the spin polarization at $E_F$ of ferromagnetic atoms at CoFe|X interfaces, affecting the ferromagnetic layer further. As a result, conductance of Hf-capped MTJ in APC is large while that of W-capped MTJ is small, explaining the difference in TMR ratios further. Note that in experiments, the MTJ transport condition is not ideally ballistic due to the amorphous lattice, the interface roughness and atoms diffusion. Therefore, the TMR ratios might be damaged.[32,33]

In summary, we investigated the spin-resolved conductance and TMR ratio in X|CoFe|MgO|CoFe|X MTJ with capping material X of W, Ta or Hf. It has been shown that TMR ratios are sensitive to different capping materials and tungsten is an ideal candidate material to obtain a giant TMR ratio. The spin-resolved transport properties were investigated in BZ. We found that in PC, the majority-spin transport channel shows a broad peak centered at $\mathbf{k}_\parallel = (0,0)$ appears due to the $\Delta_1$ state, while in the minority-spin channel the resonant transmission happens at some special $\mathbf{k}_\parallel$-points called hot spots which greatly contributes to the minority-spin conductance. The layer-resolved DOS shows the existence of interfacial resonant states. We focused on the DOSS analysis from the aspect of band symmetry, further explaining the influence on conductance and TMR ratios caused by different capping layers. This work benefits the design and optimization of MTJs by revealing the influence on TMR caused by capping layers in CoFe|MgO|CoFe MTJs.

The authors gratefully acknowledge the International Collaboration Project 2015DFE12880 and B16001, National Natural Science Foundation of China (Grant No. 61571023, 61501013, 61627813), and Beijing Municipal of Science and Technology (Grant No. D15110300320000) for their financial support of this work. We thank Prof. Hong Guo and Dr. Yibin Hu for fruitful discussions. Jiaqi Zhou acknowledges the support from China Scholarship Council (CSC) program.

[1] M. Julliere, Phys. Lett. A **54**, 225 (1975).

[2] C. Chappert, A. Fert, and F. N. Van Dau, Nature Mater. **6**, 813 (2007).

[3] J. Zhu and C. Park, Mater. Today **9**, 36 (2006).

[4] S. Yuasa, T. Nagahama, A. Fukushima, Y. Suzuki, and K. Ando, Nature Mater. **3**, 868 (2004).

[5] S. S. P. Parkin, C. Kaiser, A. Panchula, P. M. Rice, B. Hughes, M. Samant, and S.-H. Yang, Nature Mater. **3**, 862 (2004).

[6] Y. M. Lee, J. Hayakawa, S. Ikeda, F. Matsukura, and H. Ohno, Appl. Phys. Lett. **90**, 232510 (2007).

[7] S. Ikeda, J. Hayakawa, Y. Ashizawa, Y. M. Lee, K. Miura, H. Hasegawa, M. Tsunoda, F. Matsukura, and H. Ohno, Appl. Phys. Lett. **297**, 082508 (2008).

[8] J. Mathon and a. Umerski, Phys. Rev. B **63**, 220403(R) (2001).

[9] A. D. Kent and D. C. Worledge, Nature Nanotech. **10**, 187 (2015).

[10] W. H. Butler, X. G. Zhang, T. C. Schulthess, and J. M. MacLaren, Phys. Rev. B **63**, 054416 (2001).

[11] X. G. Zhang and W. H. Butler, Phys. Rev. B **70**, 172407 (2004).

[12] D. Waldron, V. Timoshevskii, Y. Hu, K. Xia, and H. Guo, Phys. Rev. Lett. **97**, 226802 (2006).

[13] Y. Ke, K. Xia, and H. Guo, Phys. Rev. Lett. **105**, 236801 (2010).

[14] V. Timoshevskii, Y. Hu, É. Marcotte, and H. Guo, J. Phys.: Condens. Matter **26**, 015002 (2014).

[15] B. S. Tao, H. X. Yang, Y. L. Zuo, X. Devaux, G. Lengaigne, M. Hehn, D. Lacour, S. Andrieu, M. Chshiev, T. Hauet, F. Montaigne, S. Mangin, X. F. Han, and Y. Lu, Phys. Rev. Lett. **115**, 157204 (2015).

[16] S. Ikeda, K. Miura, H. Yamamoto, K. Mizunuma, H. D. Gan, M. Endo, S. Kanai, J. Hayakawa, F. Matsukura, and H. Ohno, Nature Mater. **9**, 721 (2010).

[17] W.-G. Wang, M. Li, S. Hageman, and C. L. Chien, Nature Mater. **11**, 64 (2011).

[18] Z. Wang, M. Saito, K. P. McKenna, S. Fukami, H. Sato, S. Ikeda, H. Ohno, and Y. Ikuhara, Nano Lett. **16**, 1530 (2016).

[19] P. G. Gowtham, G. M. Stiehl, D. C. Ralph, and R. a. Buhrman, Phys. Rev. B **93**, 024404 (2016).

[20] G.-G. An, J.-B. Lee, S.-M. Yang, J.-H. Kim, W.-S. Chung, and J.-P. Hong, Acta Mater. **87**, 259 (2015).

[21] S. Peng, M. Wang, H. Yang, L. Zeng, J. Nan, J. Zhou, Y. Zhang, A. Hallal, M. Chshiev, K. L. Wang, Q. Zhang, and W. Zhao, Sci. Rep. **5**, 18173 (2015).

[22] Y.-W. Oh, K.-D. Lee, J.-R. Jeong, and B.-G. Park, J. Appl. Phys. **115**, 17C724 (2014).

[23] T. Liu, J. W. Cai, and L. Sun, AIP Advances **2**, 032151 (2012).

[24] S.-E. Lee, T.-H. Shim, and J.-G. Park, NPG Asia Mater. **8**, e324(2016).

[25] S.-E. Lee, Y. Takemura, and J.-G. Park, Appl. Phys. Lett. **109**, 182405 (2016).

[26] W. Wang (private communication, 2016).

[27] G. Kresse and J. Furthmüller, Phys. Rev. B **54**, 11169 (1996).

[28] J. Taylor, H. Guo, and J. Wang, Phys. Rev. B **63**, 245470 (2001).

[29] J. Taylor, H. Guo, and J. Wang, Phys. Rev. B **63**, 121104 (2001).

[30] S. Yuasa, T. Nagahama, and Y. Suzuki, Science **297**, 234 (2002).

[31] B. Wang, J. Li, Y. Yu, Y. Wei, J. Wang, and H. Guo, Nanoscale **8**, 3432 (2016).

[32] L. L. Tao, S. H. Liang, D. P. Liu, H. X. Wei, J. Wang, and X. F. Han, Appl. Phys. Lett. **104**, 172406 (2014).

[33] N. Miyakawa, D. C. Worledge, and K. Kita, IEEE Trans. Magn. Lett. **4**, 1000104(2013).

# Figures

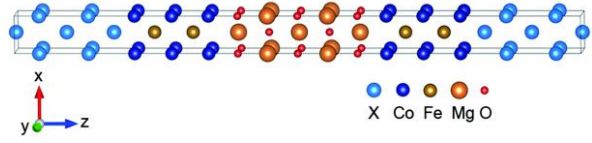

FIG. 1. Atomic structure of the X|CoFe|MgO|CoFe|X MTJ model, X represents the capping layer material W, Ta or Hf. The transport direction is along the z-axis while the MTJ is periodically repeated along the x- and y- directions.

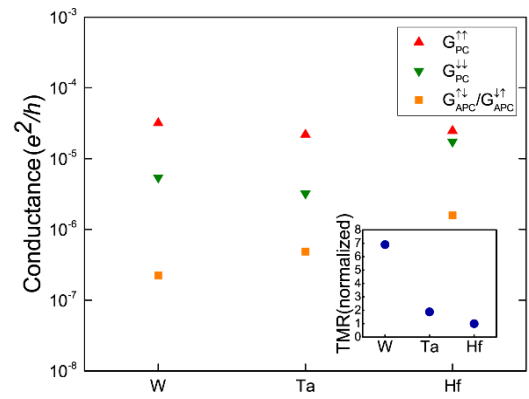

FIG. 2. Spin-resolved conductance in X-capped MTJs. X represents W, Ta or Hf as shown in the horizontal axis. The conductance unit is $e^2/h$. Inset: TMR ratios of X-capped MTJs. TMR ratios are normalized by TMR in Hf-capped MTJ for the sake of comparison.

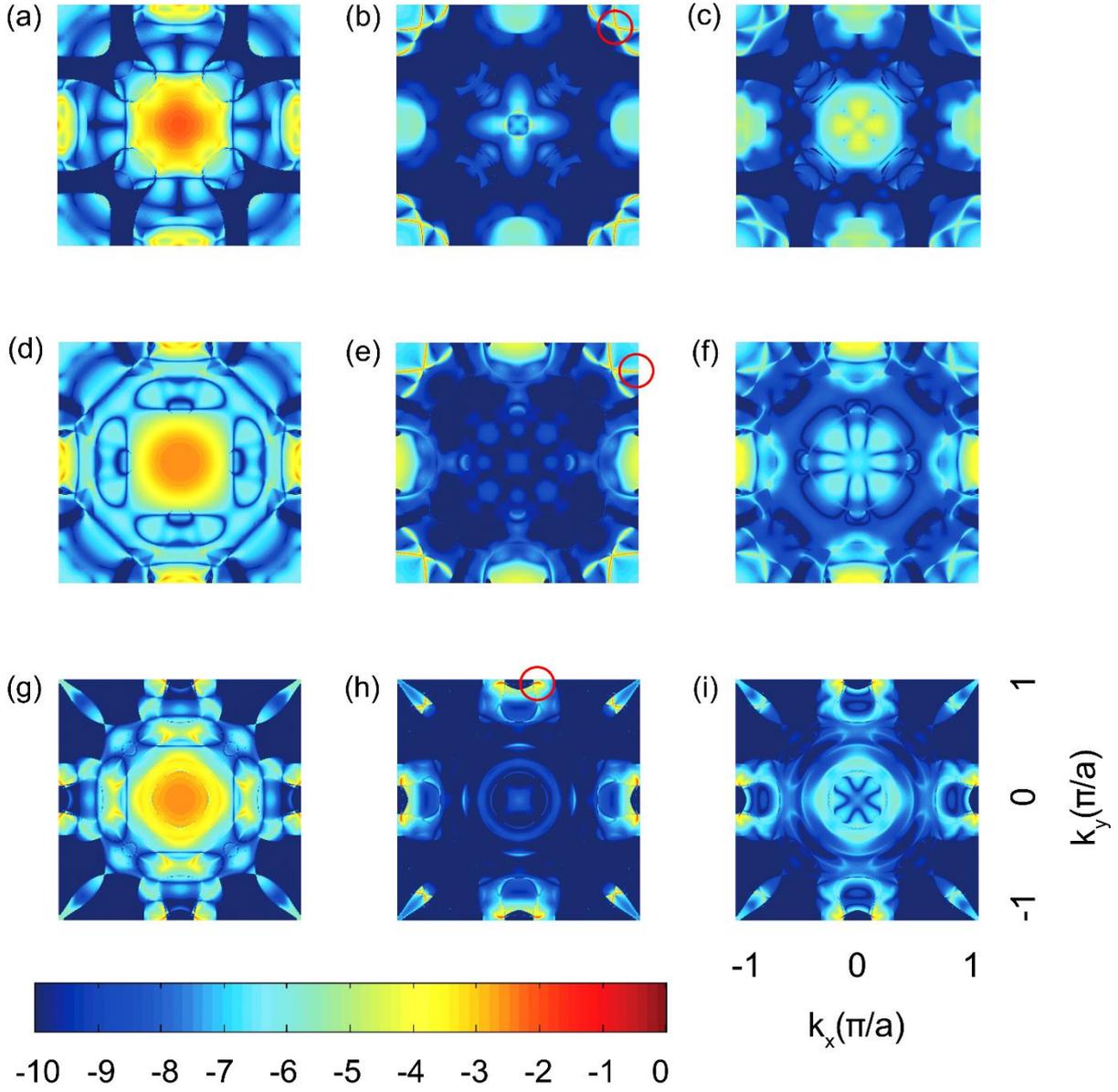

FIG. 3. Spin- and $\mathbf{k}_\parallel$-resolved transmission coefficients for ((a)-(c)) W-capped MTJ, ((d)-(f)) Ta-capped MTJ and ((g)-(i)) Hf-capped MTJ at $E_F$. Panels from left to right are ((a), (d), (g)) for majority-to-majority in PC; ((b), (e), (h)) for minority-to-minority in PC and ((c), (f), (i)) for majority-to-minority or minority-to-majority in APC. Resonant tunnel transmission features are shown in red circles.

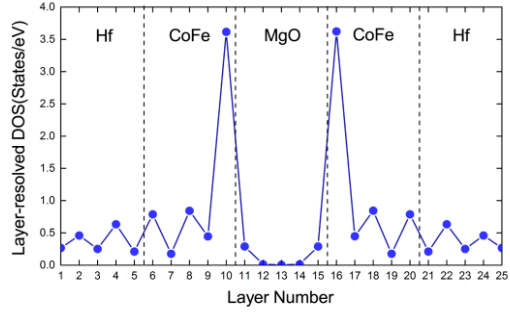

FIG. 4. Layer-resolved density of minority-spin states at $\mathbf{k}_\parallel = (0.12, 0.97)$ $\pi/a$ point for Hf-capped MTJ. Density peaks appear at CoFe|MgO interfaces, contribute to the resonant tunnel transport.

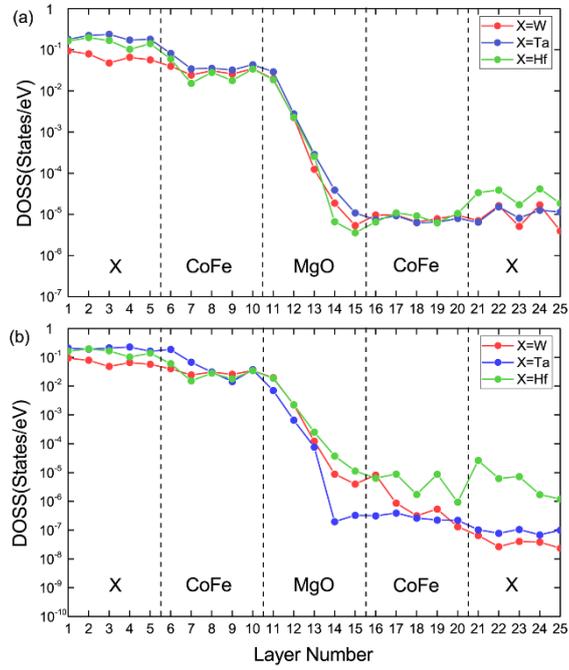

FIG. 5. Density of Scattering States (DOSS) of $\Delta_1$ states on each atomic layer for X-capped MTJ. (a) For majority- to majority-spin channel in PC; (b) for majority- to minority-spin channel in APC. X represents W, Ta or Hf.